\documentclass[11pt]{article}
\usepackage[hmargin=1in,vmargin=1in]{geometry}
\usepackage[dvipsnames,usenames,table]{xcolor}
\usepackage{hyperref}
\hypersetup{colorlinks=true, urlcolor=Blue, citecolor=Green, linkcolor=BrickRed, breaklinks, unicode}

\usepackage{hchang}

\newtheorem{theorem}{Theorem}[section]
\newtheorem{lemma}[theorem]{Lemma}
\newtheorem{claim}[theorem]{Claim}

\newtheorem{remark}[theorem]{Remark}

\newtheorem{definition}[theorem]{Definition}
\newtheorem{question}[theorem]{Question}

\newcommand{\len}[1]{{\norm{#1}}}
\def\eps{\e}
\newcommand{\weight}{w}
\DeclareMathOperator\dist{\delta}

\DeclareMathOperator{\level}{\mathit{lv}}

\begin{document}

\begin{titlepage}
	
\title{Resolving the Steiner Point Removal Problem \\in Planar Graphs via Shortcut Partitions}

\author{%
Hsien-Chih Chang%
\thanks{Department of Computer Science, Dartmouth College. Email: {\tt hsien-chih.chang@dartmouth.edu}.}  
\and 
Jonathan Conroy%
\thanks{Department of Computer Science, Dartmouth College. Email: {\tt jonathan.conroy.gr@dartmouth.edu}}  
\and 
Hung Le%
\thanks{Manning CICS, UMass Amherst. Email: {\tt hungle@cs.umass.edu}}  
\and
Lazar Milenkovi\'{c}%
\thanks{Tel Aviv University. Email: {\tt lazarm@mail.tau.ac.il}}  
\and
Shay Solomon%
\thanks{Tel Aviv University. Email: {\tt shayso@tauex.tau.ac.il}}  
\and
Cuong Than%
\thanks{Manning CICS, UMass Amherst. Email: {\tt cthan@cs.umass.edu}}  
}

\date{June 9, 2023}

\maketitle
	
\thispagestyle{empty}
	
\begin{abstract}
Recently the authors~\cite{CCLMST23} introduced the notion of \emph{shortcut partition} of planar graphs and obtained several results from the partition, including a \emph{tree cover} with $O(1)$ trees for planar metrics and an \emph{additive embedding} into small treewidth graphs. In this note, we apply the same partition to resolve the \emph{Steiner point removal (SPR)} problem in planar graphs:  Given any set $K$ of \emph{terminals} in an arbitrary edge-weighted planar graph $G$, we construct a minor $M$ of $G$ whose vertex set is~$K$, which preserves the shortest-path distances between all pairs of terminals in $G$ up to a \emph{constant} factor. This resolves in the affirmative an open problem that has been asked repeatedly in literature.
\end{abstract}

\end{titlepage}

\section{Introduction}

In the \EMPH{Steiner Point Removal} (\EMPH{SPR}) problem, we are given an undirected weighted graph $G = (V,E,\weight)$ with vertex set $V$, edge set $E$,  nonnegative weight function $\weight$ over the edges, and a subset $K$ of $V$. The vertices in $K$ are called \EMPH{terminals} and the vertices in $V \setminus K$ are called \EMPH{non-terminal} or \EMPH{Steiner} vertices. 
The goal in the SPR problem is to find a graph {minor} $M$ of $G$ such that $V(M) = K$, and for every pair $t_1, t_2$ 
of terminals in $K$, $\dist_M(t_1, t_2) \leq \alpha \cdot\dist_G(t_1, t_2)$, for some {constant} $\alpha \geq 1$; such a graph minor $M$ of $G$ is called a \EMPH{distance preserving minor} of $G$ with \EMPH{distortion~$\alpha$}. 

Gupta~\cite{Gupta01} was the first to consider the problem of removing Steiner points to preserve all terminal distances on (weighted) \emph{trees} with distortion $8$. Chan \etal~\cite{CXKR06} observed that Gupta's construction  produces a
distance preserving minor of the input tree with distortion 8, and showed a matching lower bound: There exists a tree and a set of terminals, such that any distance preserving minor of that tree must have distortion at least $8(1-o(1))$. They posed the following question: 
\begin{question} \label{q:main}
Does every planar graph admit a distance preserving minor with a \emph{constant distortion}?
\end{question} 
Question~\ref{q:main} has attracted significant research attention over the years, and numerous works have attempted to attack it from different angles. 
Some works have introduced new frameworks~\cite{Filtser18,FKT19,Filtser20} that simplify known results, others considered the problem for general graphs, establishing the distortion bound of $O(\log k)$~\cite{EGKRTT14,KNZ15,Cheung2018,Filtser18}, and there are also variants where Steiner points are allowed but their number should be minimized~\cite{KNZ14,CGH16}. 
Recently, Hershkowitz and Li~\cite{HL22} provided a solution for the SPR problem in \emph{series-parallel graphs}, extending an earlier result by Basu and Gupta~\cite{BG08} that provided a solution for \emph{outerplanar graphs}. Both outerplanar and series-parallel graphs are very restricted classes of planar graphs. For slightly larger graph classes, such as treewidth-$3$ planar graphs or $k$-outerplanar graphs for any constant $k$, the SPR problem has remained open to  date.  

In this note we resolve 
Question~\ref{q:main} in the affirmative, thus solving the SPR problem for planar graphs in its full generality:
\begin{theorem}
    \label{thm:spr-sol}
    Let $G = (V, E, \weight)$ be an arbitrary edge-weighted planar graph and let $K \subseteq V$ be an arbitrary set of terminals. Then, there is a solution for the SPR problem on $G$ with distortion $O(1)$.
\end{theorem}

The notion of \EMPH{shortcut partition} of planar graphs, introduced very recently by the authors~\cite{CCLMST23},
is the key in the proof of Theorem~\ref{thm:spr-sol}. More specifically, we apply shortcut partitions on top of the framework introduced by Filtser~\cite{Filtser20B} to obtain our result. We note that the constant in the distortion in Theorem~\ref{thm:spr-sol} is rather large; for the simplicity of presentation, we do not attempt to minimize the constant. It is an interesting open problem to reduce the distortion to a more practical constant. 

\section{Steiner Point Removal for planar graphs}

Filtser~\cite{Filtser20B} presented a reduction from the SPR problem to that of finding \emph{scattering partitions}. 
To prove Theorem~\ref{thm:spr-sol}, we introduce the notion of \emph{approximate} scattering partition (refer to Definition~\ref{def:scatter}),
and adapt the reduction of \cite[Theorem~1]{Filtser20B} using that notion. 
The first challenge underlying this adaptation is that, unlike shortest paths, an approximate shortest path does not have the optimal substructure property (any subpath of a shortest path is also a shortest path).  
The second and perhaps more significant challenge stems from the fact that we employ an inherently relaxed notion of approximate shortest paths, as we detail below.
Consequently, we have to make two crucial changes in the reduction, and change various parts of the analysis; we will point out specific changes along the way.

\begin{definition}[Approximate Scattering Partition] \label{def:scatter}
    Let $G = (V, E, \weight)$ be an edge-weighted graph. A \EMPH{$\beta$-approximate $(\tau, \Delta)$-scattering partition} of $G$ is a partition $\mathcal{C}$ of $V$ such that: 
    \begin{itemize}
        \item \textnormal{[Diameter.]} For each cluster $C$ in $\mathcal{C}$, the induced subgraph $G[C]$ has weak diameter at most $\Delta$; that is, $\dist_{G}(u,v) \le \Delta$ for any vertices $u$ and $v$ in $C$.
        \item \textnormal{[Scattering.]} For any two vertices $u$ and $v$ in $V$ such that $\dist_G(u, v) \leq \Delta$, there exists a path $\pi$ in $G$ between $u$ and $v$ where (1) $\pi$ has length at most $\beta \cdot \Delta$, (2) every edge in $\pi$ has length at most $\Delta$, and (3) $\pi$ intersects at most $\tau$ clusters in $\mathcal{C}$.
        We say $\pi$ is a \EMPH{$\beta$-approximate $(\tau,\Delta)$-scattered path}.
    \end{itemize}
\end{definition}
We remark that scattering properties (2) and (3) together imply property (1):  the length of $\pi$ is at most $O(\tau)\cdot\Delta$. Nevertheless, we prefer to keep property (1) separately from properties (2) and (3) in the definition to emphasize the fact that $\pi$ is an approximate path.  
 
Notice that the notion of approximate scattering partition is more relaxed than the original notion of scattering partition \cite{Filtser20B}. 
A scattering partition requires \emph{every} shortest path with length at most $\Delta$ to be $\tau$-scattered.
However in an approximate scattering partition there are three relaxations: 
\begin{enumerate}
\item we only require that \emph{one} such path exists; 
\item that path may be an approximate shortest path (rather than an exact shortest path);
\item the $\beta$-approximation to the length of such path $\pi$ is \emph{not} with respect to the distance between the endpoints;
rather, the length of $\pi$ is bounded by $\beta$ times $\Delta$, the diameter bound of clusters.  
\end{enumerate}
The following lemma, 
which we prove in \S2.1 and \S2.2, is analogous to Theorem 1 by Filtser~\cite{Filtser20B}, except for the key difference that we employ approximate scattering partitions. 
It implies that, somewhat surprisingly,
despite the three aforementioned relaxations introduced by our notion of approximate scattering partitions --- especially the third one that significantly relaxes the meaning of $\beta$-approximation --- such partitions still suffice for solving the SPR problem.

\begin{lemma}
    \label{lem:scattering-spr}
    Let $G$ be a graph such that for every $\Delta > 0$, every induced subgraph of $G$ admits a $\beta$-approximate $(\tau, \Delta)$-scattering partition, for some constants $\beta, \tau \geq 1$. Then, there is a solution to the SPR problem on $G$ with distortion $O(\tau^{8} \cdot \beta^5) = O(1)$.
\end{lemma}

To construct approximate scattering partitions, we use \EMPH{shortcut partitions} recently introduced by~\cite{CCLMST23} and proven to exist for planar graphs,
which we restated here.
A \EMPH{clustering} of a graph $G$ is a partition of the vertices into clusters $\mathcal{C} = \set{C_1, \ldots, C_m}$, such that each cluster $C$ induces a connected subgraph in $G$.
The \EMPH{cluster graph} of $G$ with respect to $\mathcal{C}$, denoted \EMPH{$\check{G}$}, is the graph obtained by contracting each cluster in $\mathcal{C}$ into a \emph{supernode}. The \EMPH{hop-length} of a path is the number of edges in the path.
\begin{lemma}[Corollary~3.12 in \cite{CCLMST23}]
\label{lem:cclmst-shortcut}
    For any planar graph $G$ and any $\e > 0$, there is a clustering $\mathcal{C}$ of $G$ into clusters of strong diameter at most $\e \cdot \diam(G)$ with the following property:
    For any vertices $u$ and $v$ in $G$, there is a path $\check{\pi}$ in the cluster graph $\check{G}$ between the clusters containing $u$ and $v$ with hop-length at most $O \left( \frac{\dist_G(u,v)}{\e \cdot \diam(G)} \right)$.%
    \footnote{We remark that~\cite{CCLMST23} proved something stronger: Their Corollary 3.12 further guarantees that there exists a path $\pi$ between $u$ and $v$ of length $O(1) \cdot \dist_G(u,v)$, such that $\check{\pi}$ (as defined above) only uses a subset of clusters that $\pi$ intersects. For our purposes, this added condition will not be relevant, so we omit it from the statement of Lemma~\ref{lem:cclmst-shortcut}.}
\end{lemma}

We next show that the existence of an approximate scattering partition follows from the existence of a shortcut partition.

\begin{lemma} \label{lem:scatter}
    There are constants $\beta$ and $\tau$ such that, for any planar graph $G$ and any $\Delta > 0$, there exists a $\beta$-approximate $(\tau, \Delta)$-scattering partition of $G$.
\end{lemma}
\begin{proof}
    Construct graph $G'$ from $G$ by removing all edges of length greater than $\Delta$. 
    Notice that if any pair $u,v$ of vertices satisfies $\dist_G(u,v) \le \Delta$, then it also satisfies $\dist_{G'}(u,v) \le \Delta$. Thus, any partition of vertices that satisfies the approximate scattering property for $G'$ also satisfies that property for $G$.

    Let $\mathcal{C}$ be the clustering guaranteed by Lemma~\ref{lem:cclmst-shortcut} for the graph $G'$, 
    with parameter $\e \coloneqq \Delta/\diam(G')$.
    Notice that $\mathcal{C}$ is a clustering of the vertices of $G$, where for any cluster $C$ in $\mathcal{C}$, the induced subgraphs $G[C]$ and $G'[C]$ have strong diameter at most $\eps \cdot \diam(G') = \Delta$; thus, $\mathcal{C}$ satisfies the diameter property of approximate scattering partition. 
    
    We now show that $\mathcal{C}$ satisfies the scattering property. Let $u$ and $v$ be two vertices in $G$ with $\dist_G(u,v) \le \Delta$.
    Note that $\dist_{G'}(u,v) \le \Delta$.
    By Lemma~\ref{lem:cclmst-shortcut}, there is a path $\check{\pi}$ in the cluster graph $\check{G}$ between the clusters containing $u$ and $v$, such that $\check{\pi}$ has hop-length 
    $O \left( \frac{\dist_{G'}(u,v)}{\e \cdot \diam(G')} \right) = O(1)$.
    In other words, the hop-length of $\check{\pi}$ is
    $t$, for some $t$ that is upper-bounded by a universal constant $\tau$. Write $\check{\pi} = (C_1, C_2, \ldots, C_t)$ as a sequence of $t$ adjacent clusters in $\check{G}$. Notice that two clusters $C$ and $C'$ in $\check{G}$ are adjacent if and only if there is an edge in $G'$ between a vertex in $C$ and a vertex in $C'$.
    For every pair of consecutive clusters $C_i$ and $C_{i+1}$ in $\check{\pi}$, let $x_{i}'$ be a vertex in $C_i$ and $x_{i+1}$ be a vertex in $C_{i+1}$ such that there is an edge $e_i$ in $G'$ between $x_{i}'$ and $x_{i+1}$. To simplify notation, define $x_1 \coloneqq u$ and define $x_t' \coloneqq v$. With this definition, $x_i$ and $x_i'$ are defined for all $i$ in $\set{1, \ldots, t}$.
    Notice that for every $i$ in $\set{1, \ldots, t}$, vertices $x_i$ and $x_i'$ are both in cluster $C_i$. By the strong diameter property of $\mathcal{C}$ guaranteed by~Lemma~\ref{lem:cclmst-shortcut}, there is a path $P_i$ in $G'$ between $x_i$ and $x_i'$, such that $P_i$ is contained in $C_i$ and has length at most $\Delta$.

    We define the path $\pi$ in $G'$ (and thus also in $G$) between $u$ and $v$ to be the concatenation $P_1 \circ e_1 \circ P_2 \circ e_2 \circ \ldots \circ P_t$. Notice that (1) $\pi$ has length at most $2 t \cdot \Delta \leq 2 \tau \cdot \Delta$; indeed, each subpath $P_i$ has length at most $\Delta$ (by the strong diameter property), and each edge $e_i$ has length at most $\Delta$ (as $e_i$ is in $G'$). Further, (2) every edge of $\pi$ has length at most $\Delta$, and (3) $\pi$ intersects at most $\tau$ clusters (namely, the clusters $C_1, \ldots, C_t$ along $\check{\pi}$).
\end{proof}

Combining Lemma~\ref{lem:scattering-spr} and Lemma~\ref{lem:scatter} proves Theorem~\ref{thm:spr-sol}.
In what follows we prove Lemma~\ref{lem:scattering-spr}.

\subsection{Algorithm}
Our construction for proving Lemma~\ref{lem:scattering-spr} is similar to that of 
\cite{Filtser20B}, but deviates from it in several crucial points (see Remark~\ref{rem:diff} for details).
For completeness, we next provide the entire construction of
\cite{Filtser20B}, adapted appropriately to our purposes.

We will assume without loss of generality that the minimum pairwise distance is 1.
We shall partition $V$ into $|K|$ connected subgraphs,  
each of which corresponds to a single terminal in $K$. 
Each vertex in $V$ will be \emph{assigned} to a connected subgraph by the \EMPH{assignment function $f: V \rightarrow K$}, 
such that at the end of the process, 
we can create a graph minor $M$ of $G$ by contracting each connected subgraph $f^{-1}(t)$ into a supernode for every terminal $t \in K$. 
By setting $\weight_M(t, t') \coloneqq \dist_G(t, t')$ for each edge $(t, t') \in E(M)$, 
the edge-weighted graph $M = (K, E(M), \weight_M)$ is our solution to the SPR problem on $G$. For a path $P$, we denote by $||P||$ the length of $P$.

We compute the assignment function $f$ in iterations.
In iteration $i$ we shall compute a function $\EMPH{$f_i$}: V \rightarrow K \cup \{\perp\}$, where $\perp$ symbolizes that the vertex remains unassigned.
The function $f$ will be obtained as the function $f_i$ computed at the last iteration of the algorithm.
We will maintain the set of \EMPH{relevant vertices} $\EMPH{$\mathcal{R}_i$} \coloneqq \Set{ v \in V \mid \zeta^{i - 1} \leq \dist_G(v, K) < \zeta^i}$ and the set of \EMPH{assigned vertices $V_i$} by the function $f_i$ to some terminals, for each iteration $i$, where $\EMPH{$\zeta$} \coloneqq c \cdot \beta \cdot \tau$, for $\beta$ and $\tau$ being the constants provided by Lemma~\ref{lem:scatter} and $c$ being some large constant. 
Initialize $f_0(t) \coloneqq t$ for each $t \in K$, and $f_0(v) \coloneqq \,{\perp}$ for each $v \in V \setminus K$.
Define both \EMPH{$\mathcal{R}_0$} and \EMPH{$V_0$} to be $K$.
Inductively, we maintain the properties that $V_{i-1} \subseteq V_{i}$ and $\bigcup_{j \le i}
\mathcal{R}_j \subseteq V_i$,
hence the algorithm terminates when all vertices have been assigned.

At the $i$-th iteration of the algorithm, we compute $\beta$-approximate $(\tau, \zeta^{i-1})$-scattering partition~\EMPH{$\mathcal{P}_i$}, 
provided by Lemma~\ref{lem:scatter}, on the subgraph induced on the unassigned vertices $\EMPH{$G_i$} \coloneqq G[V \setminus V_{i-1}]$.
Let~\EMPH{$\mathcal{C}_i$} be the set of clusters in $\mathcal{P}_i$ that contain at least one vertex in $\mathcal{R}_i$. 
All vertices in the clusters of $\mathcal{C}_i$ will be assigned by $f_i$ at iteration $i$. 

We classify the clusters in $\mathcal{C}_i$
into \EMPH{levels}, starting from level 0, viewing $V_{i-1}$ as a \emph{level-0} cluster. 
We say that a cluster $C \in \mathcal{C}_i$ is at \EMPH{level $j$} if $j$ is the minimum index such that there is an edge of weight at most $\zeta^i$  connecting a vertex $u$ in $C$ and another vertex $v$ in some level-$(j-1)$ cluster $C'$.
If there are multiple such edges, we fix one of them arbitrarily;
we call vertex $v$ in $C'$ the \EMPH{linking vertex} of $C$.  
Let \EMPH{$\level_i(C)$} denote the level of $C$. 
Observe that every $\mathcal{C}_i$ contains a vertex in $\mathcal{R}_i$, i.e., there exists a vertex $v \in \mathcal{C}_i$ such that $\zeta^{i - 1} \leq \dist_G(v, K) < \zeta^i$. Hence, it is readily verified that every cluster $\mathcal{C}_i$ has a linking vertex,
and thus $\EMPH{$\level_i(C)$}$ is a valid level.

For every vertex $v \in V_{i-1}$, we set $f_{i}(v) \coloneqq f_{i-1}(v)$.
For every vertex not in $\bigcup \mathcal{C}_i$ (or $V_{i-1}$), we set $f_{i}(v) = \perp$.
Next, we scan all clusters in $\mathcal{C}_i$ by non-decreasing order of level, starting from level 1. 
For each vertex $u$ in each cluster $C$, 
we set $f_i(u)$ to be $f_i(v_C)$, where $v_C$ is the linking vertex of $C$.
If some unassigned vertices remain, we proceed to the next iteration; otherwise, the algorithm terminates.

\begin{remark} 
\label{rem:diff}
The algorithm presented here is different than that of \cite{Filtser20B} in two crucial points: 
\begin{itemize}
    \item First, as mentioned, we use approximate scattering partitions (as in Definition~\ref{def:scatter}) rather than the scattering partitions of \cite{Filtser20B}.  This change poses several technical challenges in the argument.
    \item To cope with approximate scattering partitions, we do not use constant 2 as in \cite{Filtser20B} but rather use a bigger constant $\zeta$ (as defined above).
\end{itemize}
\end{remark}

\subsection{Distortion Analysis}
From the algorithm, any vertex within distance between $\zeta^{i-1}$ to $\zeta^i$ from $K$ is assigned at iteration at most~$i$.
However, the following claim narrows the possibilities down to two choices. The claim is analogous to Claim 5 in \cite{Filtser20B}, where we use $\zeta$ instead of 2, and its proof is similar.

\begin{claim}[{\cite[Claim~5]{Filtser20B}}]
    \label{clm:assign-itr}
    Any vertex $v$ satisfying $\zeta^{i - 1} \le \dist_G(v, K) < \zeta^i$ is assigned during iteration $i - 1$ or $i$.  
    Consequently, any vertex $v$ assigned during iteration $i$ must satisfy $\zeta^{i - 1} \le \dist_G(v, K) < \zeta^{i+1}$.
\end{claim}
\begin{proof}
If $v$ remains unassigned until iteration $i$, it will be assigned during iteration $i$ by construction.
Suppose that $v$ was assigned during iteration $j$. Then $v$ belongs to a cluster $C \in \mathcal{C}_j$, and there
is a vertex $u \in C$ with $\dist_G(u,K) \le \zeta^j$. As $C$ has strong diameter at most $\zeta^{j-1}$ and $\zeta > 2$, we obtain 
\[
\zeta^{i-1} \le \dist_G(v,K) \le
\dist_G(u,K) + \dist_G(u,v) \le \zeta^j + \zeta^{j-1} < \zeta^{2} \cdot \zeta^{j-1},
\]
implying that $i-1 < 2 + (j-1)$, or equivalently
$j \ge i-1$.
\end{proof}

\noindent The following claim is analogous to Corollary 1 in \cite{Filtser20B}, but we introduce a few changes in the proof.

\begin{claim}[{\cite[Corollary~1]{Filtser20B}}]
    \label{clm:real-dist}
    For every $v \in V$, $\dist_G(v, f(v)) \leq 3\tau \cdot \zeta^2 \cdot \dist_G(v, K)$.
\end{claim}

\begin{proof}
Let $i$ be the iteration in which $v$ is assigned, and let \EMPH{$C_v$} be the cluster in $\mathcal{C}_i$ containing $v$.     
We shall prove that 
\begin{equation}
\begin{aligned}
\label{eq:forind}
\dist_G(v, f(v)) \leq 3 \tau \cdot \zeta^{i+1}.
\end{aligned}
\end{equation}
Combining this bound with Claim~\ref{clm:assign-itr} yields 
\[
\dist(v, f(v)) \leq 3 \tau \cdot \zeta^{i+1} \leq 3 \tau \cdot \zeta^2 \cdot \dist_G(v, K),
\] 
as required.
The proof is by induction on the iteration $i$ in which $v$ is assigned. 
The base case  $i = 0$ is trivial, as then $v$ is a terminal, and we have $\dist_G(v, f(v)) = 0 \leq 3\tau \cdot \zeta^{0+1}$. 
We henceforth consider the induction step when $i \ge 1$.  

First, we argue that
$\level_i(C_v) \leq \zeta \cdot \tau$.
Since cluster $C_v$ is in $\mathcal{C}_i$, there exists a vertex $u \in C_v$ such that $\dist_G(u, K) < \zeta^i$. 
Let $P_u \coloneqq (u_1, u_2, \ldots u_s)$ be a shortest path from $u = u_1$ to $K$ (with $\len{P_u} < \zeta^i$), let $\ell$ be the largest index such that $u_1, u_2, \ldots u_{\ell} \in V \setminus V_{i - 1}$, and define the prefix $\EMPH{$Q$} \coloneqq (u_1, u_2, \ldots u_\ell)$ of $P_u$; note that $\ell < s$ and $u_{\ell+1} \in V_{i-1}$. 
Since $\len{Q} < \len{P_u} < \zeta^i$, 
we can greedily partition $Q$ into $\zeta' \le \zeta$ sub-paths $\EMPH{$Q_1, \ldots, Q_{\zeta'}$}$, each of length at most $\zeta^{i-1}$, connected via edges of weight less than $\zeta^i$;
that is, $Q$
is obtained as the concatenation $Q_1 \circ e_1 \circ Q_2 \circ e_2 \ldots \circ e_{\zeta' -1} \circ Q_{\zeta'}$,
where $\len{Q_j} < \zeta^{i-1}$
and $\len{e_j} < \zeta^i$ for each $j$.
Consider the $\beta$-approximate $(\tau, \zeta^{i-1})$-scattering partition $\mathcal{P}_i$ (provided by Lemma~\ref{lem:scatter}), used in the $i$th iteration, on the subgraph $G_i = G[V \setminus V_{i-1}]$ induced on the unassigned vertices.
For each $j$, 
the sub-path $Q_j$ of $Q$ is contained in $G_i$
and it satisfies $\len{Q_j} \leq \zeta^{i-1}$, thus there exists a $\beta$-approximate path \EMPH{$Q'_j$} between the endpoints of $Q_j$ that is scattered by $\tau'$ clusters, with $\tau' \leq \tau$, and each edge of $Q'_j$ is of weight at most $\zeta^{i-1}$.
The path $Q'_1 \circ e_1 \circ Q'_2 \circ e_2 \ldots \circ e_{\zeta' -1} \circ Q'_{\zeta'}$
obtained from $Q$ by replacing each sub-path $Q_j$ by its scattered path $Q'_j$, is a path from $u_1$ to $u_\ell$ intersecting at most  $\zeta \cdot \tau$ clusters in $\mathcal{C}_i$. 
Since $u_{\ell}$ is in a cluster of level $1$ (because $u_{\ell+1}$ is in $V_{i-1}$, which is of level $0$), $\level_i(C_v) \leq \zeta \cdot \tau$, as required.

\medskip
\noindent We then show that $\dist_G(v, f(v)) \leq \level_i(C_v) \cdot 2 \cdot \zeta^{i} + 3\tau \cdot \zeta^{i}$ by induction on the ($i$th-iteration) level of $C_v$.
We employ a double induction, one on the iteration $i$ and the other on the level of $C_v$; aiming to avoid confusion, we shall refer to the former as the ``outer induction'' and to the latter as the ``inner induction''.

Let \EMPH{$x$} be the linking vertex of $C_v$; in particular, we have $f(v) = f(x)$. 
Let \EMPH{$x_v$} be the vertex in $C_v$ such that $(x, x_v) \in E$ and $\weight(x, x_v) \leq \zeta^i$. 
For the basis $\level_i(C_v) = 1$ of the inner induction, $x$ is assigned during iteration $i' < i$.
By the outer induction hypothesis for iteration $i'$ (i.e., substituting $i$ with $i'$ in Eq.~\ref{eq:forind}), we obtain $\delta_G(x,f(x)) \le 3 \tau \cdot \zeta^{i' + 1} \le 3 \tau \cdot \zeta^{i}$.
By the triangle inequality and
since $\zeta > 1$:
\begin{equation}
\begin{aligned}
    \dist_G(v, f(v)) &\leq \dist_G(v, x_v) + \dist_G(x_v, x) + \dist_G(x, f(v)) \\
    &\leq \zeta^{i - 1} + \zeta^{i} + \dist_G(x, f(x)) \leq \zeta^{i - 1} + \zeta^{i} + 3\tau \cdot \zeta^i \leq 2 \cdot \zeta^{i} + 3\tau \cdot \zeta^{i}.
\end{aligned}
\end{equation}
For the inner induction step, consider the case $\level_i(C_v) > 1$. Let $C_x$ be the cluster in $\mathcal{P}_i$ containing $x$; in particular, we have $\level_i(C_x) = \level_i(C_v) - 1$. By the inner induction hypothesis on the level of $C_x$, we have $\dist(x, f(x)) \leq \level_i(C_x) \cdot 2 \cdot \zeta^{i} + 3\tau \cdot \zeta^{i}$.  
Using the triangle inequality again, we have:
\begin{equation}
\begin{aligned}
    \dist_G(v, f(v)) &\leq \dist_G(v, x_v) + \dist_G(x_v, x) + \dist_G(x, f(v)) \\
    &\leq \zeta^{i - 1} + \zeta^{i} + \dist_G(x, f(x)) \leq \zeta^{i - 1} + \zeta^{i} + \level_i(C_x) \cdot 2 \cdot \zeta^{i} + 3\tau \cdot \zeta^i \\
    &\leq 2 \cdot \zeta^{i} + (\level_i(C_v) - 1) \cdot 2 \cdot \zeta^{i} + 3\tau \cdot \zeta^{i} = \level_i(C_v) \cdot 2 \cdot \zeta^{i} + 3\tau \cdot \zeta^{i},
\end{aligned}
\end{equation}
which completes the inner induction step.

Since $\level_i(C_v) \leq \zeta \cdot \tau$ and as $\zeta > 3$, it follows that $\dist(v, f(v)) \leq 3 \tau \cdot \zeta^{i+1}$, which completes the outer induction step. The claim follows. 
\end{proof}

\noindent Now we are ready to prove Lemma~\ref{lem:scattering-spr}.

\begin{proof}[of Lemma~\ref{lem:scattering-spr}]
We prove that our algorithm returns a minor of $G$ that satisfies the SPR conditions. 
Consider an arbitrary pair of terminals $t$ and $t'$. Let $P \coloneqq (v_1, v_2, \ldots, v_{|P|})$ be a shortest path between $v_1 \coloneqq t$ and $v_{|P|} \coloneqq t'$.
For each subpath $I \coloneqq (v_\ell, v_{\ell + 1}, \ldots v_r)$ of $P$, let \EMPH{$I^+$} denote the \EMPH{extended subpath} $(v_{\ell - 1}, v_\ell, v_{\ell + 1}, \ldots v_r, v_{r + 1})$; we define $v_0 \coloneqq v_1$ and $v_{|P| + 1} \coloneqq v_{|P|}$ for technical convenience. 
Partition $P$ into a set \EMPH{$\mathcal{I}$} of subpaths called \EMPH{intervals} such that for each subpath $I \in \mathcal{I}$ between $v_\ell$ and $v_r$: 
\begin{equation}
\begin{aligned}
    \label{eq:wI}
    \len{I}  \leq \eta \cdot\dist_G(v_\ell, K) \leq \len{I^+},
\end{aligned}
\end{equation}
where $\EMPH{$\eta$} \coloneqq \frac{1}{4\zeta}$.
It is easy to verify that $\mathcal{I}$ can be constructed greedily from $P$.

Consider an arbitrary interval $I = (v_\ell, v_{\ell + 1}, \ldots v_r) \in \mathcal{I}$. 
Let $u \in I$ be a vertex that is assigned in iteration $i$, and assume no vertex of $I$ was assigned prior to iteration $i$.
Since $u$ is assigned in iteration $i$, $u$ belongs to a cluster $C$ in $\mathcal{C}_i$, which is the subset of
clusters that contain at least one vertex in $\mathcal{R}_i$, among the $\beta$-approximate $(\tau, \zeta^{i-1})$-scattering partition $\mathcal{P}_i$ computed at the $i$th iteration.
Hence, by definition, $C$ has strong diameter  at most $\zeta^{i-1}$ and there exists a vertex $u' \in C$ such that $\dist_G(u', K) < \zeta^i$, implying that 
\begin{equation}
\begin{aligned}
    \label{eq:duK2}
    \dist_G(u, K) \leq 
    \dist_G(u, u') +
    \dist_{G}(u', K) 
    < 
    \zeta^{i-1} + \zeta^i
    < 
    2\zeta^{i}.   
\end{aligned}
\end{equation}
By Eq.~\ref{eq:wI} and the triangle inequality, 
\[
\dist_G(v_\ell, K) \leq \dist_G(v_\ell, u) + \dist_G(u, K) \leq \len{I} + \dist_G(u, K)
\leq \eta \cdot \dist_G(v_\ell,K) + \dist_G(u,K),
\]
which together with Eq.~\ref{eq:duK2} and the fact that $\eta < 1/2$ yields
\begin{equation}
\begin{aligned}
\label{eq:boundvlk}
    \dist_G(v_\ell,K) \le \frac{\dist_G(u,K)}{1-\eta} <  \frac{2\zeta^{i}}{1-\eta} 
    <  4 \zeta^{i}. 
\end{aligned}
\end{equation}
By Eq.~\ref{eq:wI} and Eq.~\ref{eq:boundvlk}, 
\begin{equation}
\begin{aligned}
\label{eq:distsmall}
    \dist_G(v_\ell, v_r) &= \len{I} \leq \eta \cdot \dist_G(v_\ell, K) < \eta \cdot 4\zeta^{i}  = \zeta^{i - 1},
\end{aligned}
\end{equation}
where the last inequality holds
as $\eta = \frac{1}{4\zeta}$.
    
At the beginning of iteration $i$, all vertices of $I$ are unassigned, i.e., $I$ is in $G_i = G[V \setminus V_{i-1}]$,
    and Eq.~\ref{eq:distsmall} yields
$\dist_{G_{i}}(v_\ell, v_r)
= \dist_{G}(v_\ell, v_r)
<  \zeta^{i-1}$.
At the $i$th iteration a $\beta$-approximate $(\tau, \zeta^{i - 1})$-scattering partition 
$\mathcal{P}_{i}$
on $G_{i}$ is computed, thus there exists a $\beta$-approximate 
$(\tau, \zeta^{i - 1})$-scattered
path \EMPH{$I'$} in $G_{i}$ from $v_\ell$ to $v_r$ that is scattered by at most $\tau$ clusters in $\mathcal{P}_{i}$, with $\len{I'} \le \beta \cdot \zeta^{i-1}$.
A path is called a \EMPH{detour} if its first and last vertices are assigned to the same terminal. 
Since vertices in the same cluster will be assigned to the same terminal,
at the end of iteration $i$, $I'$ can be greedily partitioned into at most $\tau$ detours and $\tau + 1$ subpaths that contain only unassigned vertices;
in other words, we can write $I' \coloneqq \EMPH{$P_1 \circ Q_1 \circ \ldots \circ P_\rho \circ Q_\rho \circ P_{\rho + 1}$}$, where $\rho \leq \tau$, $Q_1, Q_2, \ldots Q_{\rho}$ are detours, and each of the (possibly empty) sub-paths $P_1, P_2, \ldots P_{\rho + 1}$ contains only unassigned vertices at the end of iteration $i$. 

Fix an arbitrary index $j \in [1 \,..\,\rho+1]$.
Let $a_j$ and $b_j$ be the first and last vertices of $P_j$; it is possible that $a_j = b_j$.  
Since $\len{I'} \le \beta \cdot \zeta^{i-1}$ and as $\beta < \zeta$, we have
\begin{equation}
\begin{aligned} \label{eq:ajbj}
    \dist_G(a_j, b_j) \leq \len{P_j} \le \len{I'} \leq \beta \cdot \zeta^{i-1} < \zeta^i.    
\end{aligned}
\end{equation}
At the beginning of iteration $i+1$, all vertices of $P_j$ are unassigned by definition, hence $P_j$ is in $G_{i+1} = G[V \setminus V_{i}]$
and by Eq.~\ref{eq:ajbj},
$\dist_{G_{i+1}}(a_j, b_j)
\le \len{P_j} < \zeta^i$.
At the $(i+1)$th iteration a
$\beta$-approximate $(\tau, \zeta^{i})$-scattering partition  $\mathcal{P}_{i+1}$ on $G_{i+1}$ is computed, thus there exists a $\beta$-approximate 
$(\tau, \zeta^{i})$-scattered path \EMPH{$P'_j$} 
 in $G_{i+1}$ from $a_j$ to $b_j$
that is scattered by at most $\tau$
clusters in $\mathcal{P}_{i+1}$, with $\len{P'_j} \leq \beta \cdot \zeta^i$. 
    
Next, consider the path $\EMPH{$I''$} \coloneqq P'_1 \circ Q_1 \circ \ldots \circ P'_\rho \circ Q_\rho \circ P'_{\rho + 1}$. 
By Eq.~\ref{eq:distsmall} we have 
\begin{equation}
\begin{aligned} 
\label{eq:boundi''}
\len{I''} \leq 
\len{I} + \sum_{j=1}^{\rho+1} \len{P'_j} 
\le \zeta^{i-1} + (\tau+1) \beta \cdot \zeta^i 
\le (\tau+2) \beta \cdot \zeta^i
\end{aligned}
\end{equation}
Since no vertex in $I$ (in particular, $v_\ell$) was assigned prior to iteration $i$, 
Claim~\ref{clm:assign-itr} yields
$\dist_G(v_\ell,K) \ge \zeta^{i-1}$.
Eq.~\ref{eq:wI} yields $\len{I^+} \ge \eta \cdot \dist_G(v_\ell,K) \ge \eta \cdot \zeta^{i-1}$,
and as $\eta = \frac{1}{4\zeta}$ 
we obtain
\begin{equation}
\begin{aligned} 
\label{eq:idouble}
\len{I''} \leq  (\tau+2) \beta \cdot \zeta^i
\le 4 \zeta^2 (\tau+2)\beta 
\cdot \len{I^+}. 
\end{aligned}
\end{equation}
    
Next, we argue that
all vertices in $I''$ are assigned at the end of iteration $i + 1$.
Let \EMPH{$w$} be an arbitrary vertex in $I''$;
by Claim~\ref{clm:assign-itr}, it suffices to show that $\dist_G(w, K) < \zeta^{i + 1}$. 
Recall that $u$ is a vertex of $I$ that is assigned in iteration $i$.
By Eq.~\ref{eq:duK2}, Eq.~\ref{eq:distsmall}, Eq.~\ref{eq:boundi''} and the triangle inequality, 
\begin{equation}
\begin{aligned}
\label{eq:dw}
    \dist_G(w, K) &\leq \dist_G(v_\ell, K) + \dist_G(v_\ell, w) \leq \dist_G(v_\ell, u) + \dist_G(u, K) + \dist_G(v_\ell, w) \\
    &\leq \len{I} + \dist_G(u, K) + \len{I''} < \zeta^{i-1} + 2\zeta^{i} + (\tau+2) \beta \cdot \zeta^i < \zeta^{i+1}, 
\end{aligned}
\end{equation}
where the last inequality holds since
$\zeta = c \cdot \beta \cdot \tau$ for a sufficiently large constant $c$.

Hence, every vertex in $P'_j$ is assigned by iteration $i + 1$, for every $j \in [1 \,..\,\rho+1]$. Then, $P'_j$ could be greedily partitioned into at most $\tau$ detours, as before with $I'$, but we have no subpaths of unassigned vertices in $I''$, since every vertex in $I''$ must be assigned by the end of iteration $i + 1$. 
We have thus shown that $I''$ can be partitioned into at most $O(\tau^2)$ detours \EMPH{$D_1, D_2, \ldots D_g$}, with $\EMPH{$g$} = O(\tau^2)$. 
For each $j \in [1 \,..\, g]$, let \EMPH{$x_j$} and \EMPH{$y_j$} be the first and last vertices in $D_j$. 
Because $I''$ are partitioned greedily into \emph{maximal} detours, one has $f(y_j) \ne f(x_{j+1})$ for all $j$.
Observe that there exists an edge between $f(x_j)$ and $f(x_{j + 1})$ in the SPR minor $M$ for each $j \in [1 \,..\, g-1]$, since $f(x_j) = f(y_j) \in K$ and $(y_j, x_{j + 1}) \in E$.
Consequently, by the triangle inequality, Corollary~\ref{clm:real-dist} and Eq.~\ref{eq:idouble},
\begin{equation}
\begin{aligned}
\label{eq:dvlvr1}
\dist_M(f(v_\ell), f(v_r)) 
&\leq \sum_{j = 1}^{g - 1}\dist_M(f(x_j), f(x_{j + 1})) = \sum_{j = 1}^{g - 1}\dist_G(f(x_j), f(x_{j + 1})) \\
&\leq \sum_{j = 1}^{g - 1}\left[\dist_G(x_j, f(x_j)) + \dist_G(x_j, x_{j + 1}) + \dist_G(x_{j + 1}, f(x_{j + 1}))\right] \\
&\leq 2\sum_{j = 1}^{g} \dist_G(x_j, f(x_j)) + \sum_{j = 1}^{g - 1}\dist_G(x_j, x_{j + 1}) \leq 2\sum_{j = 1}^{g} \dist_G(x_j, f(x_j)) + \len{I''} \\
&\leq 6\tau \zeta^2 \sum_{j = 1}^{g} \dist_G(x_j, K) + 4 \zeta^2 (\tau+2)\beta 
\cdot \len{I^+}. 
\end{aligned}
\end{equation}
For every vertex $v'' \in I''$, we have
\begin{equation}
\begin{aligned}
\label{eq:dv-k}
    \dist_G(v'', K) &\leq \dist_G(v'', v_\ell) + \dist_G(v_\ell, K) \leq \len{I''} + \dist_G(v_\ell, K) \\
    &\leq 4 \zeta^2 (\tau+2)\beta 
    \cdot \len{I^+}  + \frac{\len{I^+}}{\eta} \leq 4 \zeta^2 (\tau+3)\beta 
    \cdot \len{I^+},
\end{aligned}
\end{equation}
where the penultimate inequality holds by
Eq.~\ref{eq:wI} and Eq.~\ref{eq:idouble} and the last inequality holds
since $\eta = \frac{1}{4\zeta}$. We remark that  Eq.~\ref{eq:dv-k} also holds for any vertex $v' \in I$, which will be used below for deriving Eq.~\ref{eq:dist-inter-interval}. Hence, for every $j \in [1 \,..\, g]$, $\dist_G(x_j, K) \leq 4 \zeta^2 (\tau+3)\beta 
\cdot \len{I^+}$; plugging this in Eq.~\ref{eq:dvlvr1} yields:
    \begin{equation}
\begin{aligned}
        \label{eq:dvlvr2}
        \dist_M(f(v_\ell), f(v_r)) \leq 24
        \zeta^4 \tau(\tau+3) \beta g \cdot \len{I^+}
        + 4 \zeta^2 (\tau+2)\beta 
\cdot \len{I^+}
         = O(\zeta^4 \cdot \tau^4 \cdot \beta)
         \cdot \len{I^+}.
    \end{aligned}
\end{equation}

Next, we bound the distance between $t$ and $t'$ in $M$. 
So far we fixed an arbitrary interval  $I = (v_\ell, v_{\ell + 1}, \ldots v_r) \in \mathcal{I}$.
Writing $\mathcal{I} = \{I_1, I_2, \ldots I_s\}$, we have 
$\sum_{j = 1}^s\len{I_j} = \len{P} = \dist_G(t,t')$, hence  
    \begin{equation}
\begin{aligned}
    \label{eq:basicsum}
        \sum_{j = 1}^s\len{I^+_j} \leq 2\len{P} = 2\cdot \dist_G(t, t'). 
    \end{aligned}
\end{equation}
For each $I_j$, let \EMPH{$v^j_\ell$} and \EMPH{$v^j_r$} be the first and last vertices of $I_j$. 
For each $j \in [1 \,..\, s-1]$, since $(v_r^j, v_\ell^{j + 1}) \in E$, either $(f(v_r^j), f(v_\ell^{j + 1})) \in E(M)$ or
$f(v_r^j) = f(v_\ell^{j + 1})$,
thus we have 
$\dist_M(f(v^j_r), f(v^{j + 1}_\ell))
= \dist_G(f(v^j_r), f(v^{j + 1}_\ell))$. Hence, using the triangle inequality:
\begin{equation}
\begin{aligned}
\label{eq:dtt'}
    \dist_M(t, t') &\leq \sum_{j = 1}^{s - 1}
    \Paren{\big. \dist_M(f(v^j_\ell), f(v^j_r)) + \dist_M(f(v^j_r), f(v^{j + 1}_\ell)) } 
    + \dist_M(f(v^s_\ell), f(v^s_r)) \\
    &\leq O(\zeta^4 \cdot \tau^4 \cdot \beta) \cdot \sum_{j = 1}^s \len{I_j^+} + \sum_{j = 1}^{s - 1}\dist_M(f(v^j_r), f(v^{j + 1}_\ell)) \qquad \text{(by Eq.~\ref{eq:dvlvr2})} \\
    &\leq O(\zeta^4 \cdot \tau^4 \cdot \beta) \cdot \dist_G(t, t') + \sum_{j = 1}^{s - 1}\dist_M(f(v^j_r), f(v^{j + 1}_\ell)).
    \qquad \text{(by Eq.~\ref{eq:basicsum})}
    \\
    &= O(\zeta^4 \cdot \tau^4 \cdot \beta) \cdot \dist_G(t, t') + \sum_{j = 1}^{s - 1}\dist_G(f(v^j_r), f(v^{j + 1}_\ell)).
\end{aligned}
\end{equation}
Using the triangle inequality again, we have: 
\begin{equation}
\begin{aligned}
\label{eq:dist-inter-interval}
    \sum_{j = 1}^{s - 1}\dist_G(f(v^j_r), f(v^{j + 1}_\ell)) &\leq \sum_{j = 1}^{s - 1}
    \Paren{\big. 
    \dist_G(f(v^j_r), v^j_r) + \dist_G(v^j_r, v^{j + 1}_\ell) + \dist_G(v^{j + 1}_\ell, f(v^{j + 1}_\ell)) } \\
    &\leq \sum_{j = 1}^{s - 1}\dist_G(v^j_r, v^{j + 1}_\ell) + \sum_{j = 1}^{s} \Paren{\big.
    \dist_G(v^j_\ell,f(v^j_\ell)) + \dist_G(v^j_r,f(v^j_r)) } \\
    &\leq \len{P} + 3\tau \zeta^2 \cdot \sum_{j = 1}^{s}
    \Paren{\big.
    \dist_G(v^j_\ell, K) + \dist_G(v^j_r, K) } \qquad \text{(by Corollary~\ref{clm:real-dist})} \\
    &\leq \dist_G(t, t') + 3\tau \zeta^2 \cdot 4 \zeta^2 (\tau+3)\beta \cdot \sum_{j = 1}^{s}(\len{I_j^+} + \len{I_j^+})  \qquad \text{(by Eq.~\ref{eq:dv-k})} \\ 
    &\leq O(\zeta^4 \cdot \tau^2 \cdot \beta) \cdot \dist_G(t, t') \qquad \text{(by Eq.~\ref{eq:basicsum})}. 
\end{aligned}
\end{equation}

\noindent Plugging Eq.~\ref{eq:dist-inter-interval} into Eq.~\ref{eq:dtt'}, we obtain $\dist_M(t, t') = O(\zeta^4 \cdot \tau^4 \cdot \beta) \cdot \dist_G(t, t')$.
Since $\zeta = O(\beta \cdot \tau)$, we conclude that 
$\dist_M(t, t') = O(\tau^8 \cdot \beta^5)$, as required.
\end{proof}


\small
\bibliographystyle{alphaurl}
\bibliography{main}

\newcommand{\etalchar}[1]{$^{#1}$}
\begin{thebibliography}{EGK{\etalchar{+}}14}

\bibitem[BG08]{BG08}
A.~Basu and A.~Gupta.
\newblock Steiner point removal in graph metrics.
\newblock Unpublished manuscript, available from
  \url{https://www.ams.jhu.edu/~abasu9/papers/SPR.pdf}, 2008.

\bibitem[CCL{\etalchar{+}}23]{CCLMST23}
Hsien-Chih Chang, Jonathan Conroy, Hung Le, Lazar Milenkovic, Shay Solomon, and
  Cuong Than.
\newblock Covering planar metrics (and beyond): O(1) trees suffice, 2023.
\newblock Arxiv.

\bibitem[CGH16]{CGH16}
Yun~Kuen Cheung, Gramoz Goranci, and Monika Henzinger.
\newblock Graph minors for preserving terminal distances approximately -- lower
  and upper bounds.
\newblock In {\em 43rd International Colloquium on Automata, Languages, and
  Programming (ICALP 2016)}, Leibniz International Proceedings in Informatics
  (LIPIcs), pages 131:1--131:14, 2016.
\newblock URL: \url{http://drops.dagstuhl.de/opus/volltexte/2016/6267}, \href
  {https://doi.org/10.4230/LIPIcs.ICALP.2016.131}
  {\path{doi:10.4230/LIPIcs.ICALP.2016.131}}.

\bibitem[Che18]{Cheung2018}
Yun~Kuen Cheung.
\newblock Steiner point removal {\textemdash} distant terminals
  don{\textquotesingle}t (really) bother.
\newblock In {\em Proceedings of the 29th Annual {ACM}-{SIAM} Symposium on
  Discrete Algorithms, (SODA `18)}, pages 1353--1360. 2018.
\newblock \href {https://doi.org/10.1137/1.9781611975031.89}
  {\path{doi:10.1137/1.9781611975031.89}}.

\bibitem[CXKR06]{CXKR06}
T.~H.~Hubert Chan, Donglin Xia, Goran Konjevod, and Andrea Richa.
\newblock A tight lower bound for the {{Steiner}} point removal problem on
  trees.
\newblock In {\em Approximation, Randomization, and Combinatorial Optimization.
  Algorithms and Techniques (APPROX/RANDOM 2006)}, pages 70--81. 2006.
\newblock \href {https://doi.org/10.1007/11830924_9}
  {\path{doi:10.1007/11830924_9}}.

\bibitem[EGK{\etalchar{+}}14]{EGKRTT14}
Matthias Englert, Anupam Gupta, Robert Krauthgamer, Harald R\"{a}cke, Inbal
  Talgam-Cohen, and Kunal Talwar.
\newblock Vertex sparsifiers: New results from old techniques.
\newblock {\em SIAM Journal on Computing}, 43(4):1239--1262, 2014.
\newblock \href {https://doi.org/10.1137/130908440}
  {\path{doi:10.1137/130908440}}.

\bibitem[Fil18]{Filtser18}
Arnold Filtser.
\newblock Steiner point removal with distortion ${O}(\log(k))$.
\newblock In {\em Proceedings of the 29th Annual {ACM}-{SIAM} Symposium on
  Discrete Algorithms, (SODA `18)}, pages 1361--1373. 2018.
\newblock \href {https://doi.org/10.1137/1.9781611975031.90}
  {\path{doi:10.1137/1.9781611975031.90}}.

\bibitem[Fil20a]{Filtser20}
Arnold Filtser.
\newblock A face cover perspective to $\ell_1$ embeddings of planar graphs.
\newblock In {\em Proceedings of the Fourteenth Annual {ACM}-{SIAM} Symposium
  on Discrete Algorithms}, pages 1945--1954. 2020.
\newblock \href {https://doi.org/10.1137/1.9781611975994.120}
  {\path{doi:10.1137/1.9781611975994.120}}.

\bibitem[Fil20b]{Filtser20B}
Arnold Filtser.
\newblock Scattering and sparse partitions, and their applications.
\newblock In {\em 47th International Colloquium on Automata, Languages, and
  Programming (ICALP 2020)}, volume 168 of {\em Leibniz International
  Proceedings in Informatics (LIPIcs)}, pages 47:1--47:20, 2020.
\newblock URL: \url{https://drops.dagstuhl.de/opus/volltexte/2020/12454}, \href
  {https://doi.org/10.4230/LIPIcs.ICALP.2020.47}
  {\path{doi:10.4230/LIPIcs.ICALP.2020.47}}.

\bibitem[FKT18]{FKT19}
Arnold Filtser, Robert Krauthgamer, and Ohad Trabelsi.
\newblock Relaxed {{Voronoi}}: A simple framework for terminal-clustering
  problems.
\newblock In {\em 2nd Symposium on Simplicity in Algorithms (SOSA 2019)}, pages
  10:1--10:14, 2018.
\newblock \href {https://doi.org/10.4230/OASIcs.SOSA.2019.10}
  {\path{doi:10.4230/OASIcs.SOSA.2019.10}}.

\bibitem[Gup01]{Gupta01}
Anupam Gupta.
\newblock Steiner points in tree metrics don't (really) help.
\newblock In {\em Proceedings of the 12th Annual ACM-SIAM Symposium on Discrete
  Algorithms}, SODA '01, page 220–227, 2001.

\bibitem[HL22]{HL22}
D.~Ellis Hershkowitz and Jason Li.
\newblock ${O}(1)$ {{Steiner}} point removal in series-parallel graphs.
\newblock In {\em 30th Annual European Symposium on Algorithms (ESA 2022)},
  pages 66:1--66:17, 2022.
\newblock \href {https://doi.org/10.4230/LIPIcs.ESA.2022.66}
  {\path{doi:10.4230/LIPIcs.ESA.2022.66}}.

\bibitem[KKN15]{KNZ15}
Lior Kamma, Robert Krauthgamer, and Huy~L. Nguy$\tilde{\hat{\mathrm{e}}}$n.
\newblock Cutting corners cheaply, or how to remove {{Steiner}} points.
\newblock {\em {SIAM} Journal on Computing}, 44(4):975--995, 2015.
\newblock \href {https://doi.org/10.1137/140951382}
  {\path{doi:10.1137/140951382}}.

\bibitem[KNZ14]{KNZ14}
Robert Krauthgamer, Huy~L. Nguy$\tilde{\hat{\mathrm{e}}}$n, and Tamar Zondiner.
\newblock Preserving terminal distances using minors.
\newblock {\em {SIAM} Journal on Discrete Mathematics}, 28(1):127--141, 2014.
\newblock \href {https://doi.org/10.1137/120888843}
  {\path{doi:10.1137/120888843}}.

\end{thebibliography}

\end{document}